\documentstyle[multicol,pre,aps,epsfig]{revtex} 
\input epsf
\newcommand{\be}{\begin{equation}}
\newcommand{\en}{\end{equation}}
\newcommand{\bea}{\begin{eqnarray}}
\newcommand{\ena}{\end{eqnarray}}

\begin{document}

\draft

\title{Optimization of congested traffic flow in systems with a localized
periodic inhomogeneity}

\author{Elad~Tomer$^1$\footnote{email: tomer@alon.cc.biu.ac.il}, 
Leonid~Safonov$^{1,2}$, Nilly~Madar$^1$, and Shlomo~Havlin$^1$}
\address{$^1$ Minerva Center and Department of Physics,
         Bar--Ilan~University,
         52900~Ramat--Gan, Israel}
\address{$^2$ Department of Applied Mathematics and Mechanics, Voronezh State
University, 394693~Voronezh, Russia}

\date{\today}

\maketitle

\begin{abstract}
We study traffic flow on roads with a localized periodic inhomogeneity such as 
traffic signals, using a stochastic car-following model. 
We find that in cases of congestion, traffic flow can be optimized by 
controlling the inhomogeneity's frequency. 
By studying the wavelength dependence of the flux in stop-and-go traffic states, and 
exploring their stability, we are able to explain the optimization process. 
A general conclusion drawn from this study is, that the fundamental 
diagram of traffic (density flux relation) has to be generalized to include 
the influence of wavelength on the flux, for the stop-and-go traffic. Projecting 
the generalized fundamental diagram on the density-flux plane yields a 
2D region, qualitatively similar to that found empirically 
[B. S. Kerner, Phys. Rev. Lett. {\bf 81}, 3797 (1998)] in synchronized flow.
\end{abstract}

\pacs{}

\begin{multicols}{2}
\noindent

The theory of traffic flow has been a subject of comprehensive study for more than 
half a century 
\cite{LW,Nagel,Wolf,ChSanSchad,Helb,Ker0,Ker1,Ker2,NSSS,THH,HR,NS,BML,Sug,Mit_Naka,TSH1,LLK,May} 
due to its theoretical and practical importance. Much attention has been devoted to 
characterizing the different states of congested traffic \cite{Ker0,Ker1,Ker2,NSSS,THH},
including synchronized flow and stop-and-go traffic.  

Optimization of systems with localized inhomogeneities such as traffic signals 
or entrance ramps was also extensively studied in the last decades 
\cite{May,Gazis,Cronje,Gartner-et-al,Zhang-et-al,ChangLin,ChowSchad}. 
Conventional traffic signals and ramp control theories are able to optimize 
such quantities as the total delay time of all drivers in the system, assuming 
a well defined flux-density relation known as the 'fundamental diagram' of traffic flow. 

This basic concept of a fundamental diagram was ingrained in traffic flow theory for
many decades. It was believed that the density-flux relation  
can be displayed as a single curve or as a combination of two isolated curves 
(see e.g. \cite{May,Ker1}). Recently, an experimental study of Kerner 
\cite{Ker1,Ker2} shows that such a fundamental diagram does not exist. 
Instead, synchronized traffic displays a two-dimensional region in the 
density-flux plane. That is, for a given value of density, there exist a range of 
possible flux values. Mechanisms to optimize traffic flow by approaching the highest 
values of this range have not yet been suggested.

In this paper we investigate whether this new insight on the nature of traffic flow
can be applied for optimizing the flux close to its maximal value for a given 
congested density. For this purpose, systems with periodic localized 
inhomogeneities are studied, using a recent car-following model with 
multiple stable and metastable states \cite{TSH1}. 
Two types of periodic inhomogeneities are considered: 
\begin{itemize}
\item[(a)] A signalized intersection, and 
\item[(b)] an on-ramp with a signalized entrance.
\end{itemize}
Our study focus on cases of oversaturation, i.e. on cases where traffic is 
congested upstream to these inhomogeneities.

In the common traffic flow microscopic models such as in Refs. 
\cite{THH,HR,NS,BML,Sug,Mit_Naka,TSH1}, the acceleration $a$ of a car depends on its 
headway $\Delta x$, velocity $v$, and velocity difference $\Delta v$ with the 
car ahead, i.e. $a=a(\Delta x,v,\Delta v)$. In particular, for the model in \cite{TSH1} 
\be
\label{eqModel}
 a =  A\left(1-{{\Delta x_0}\over{\Delta x}}\right) - 
{ {Z^2(-\Delta v)} \over {2(\Delta x - D)} } -
kZ(v-v_{per})+\eta ',
\en
where $T$ is the safety time gap, $D$ and $\Delta x_0=vT+D$ are the minimal and
the optimal distance to the car ahead. $A$, $k$, and $v_{per}$ are constants, and 
the function $Z$ is defined as $Z(x)=(x+|x|)/2$. 
In the following numerical solutions of (\ref{eqModel}), the random term $\eta'$ 
is represented by choosing a random number in every iteration, uniformly distributed in the range 
$-0.5\eta \le \eta '\le 0.5\eta$. In the following we use the same choice of 
parameters as in \cite{TSH1} with $A = 3m/sec^2$, and a numerical time interval 
$\Delta t=0.1sec$. 

{\bf (a) Signalized intersection:}
In an oversaturated signalized intersection, the congested flow in each direction 
is affected only by the parameters of the traffic signals of this direction. 
Therefore we consider a single direction for simplicity (as e.g. in \cite{Gazis}). 
To study this case, a single traffic light is 
placed at position $L/2$ on a road with length $L$ and periodic boundary conditions. 
The flux $f$ is measured at the position of the traffic light. Simulations are performed 
with different durations of the red, amber (yellow), and green lights 
($\tau_{r}$, $\tau_{y}$ \cite{amber}, $\tau_{g}$ respectively), 
and of $\tau_{-} $ which is the total time in each signal period where the 
intersection is not in use by any of its approaches. Therefore 
$\tau_{g} = \tau - \tau_{-} - (\tau_{r}+\tau_{y})$ 
where $\tau$ is the total signal period.
$P_{r} =\tau_{r} /\tau$ is the relative duration of the red light.

Our goal here is to optimize the flux in a specific road without affecting
the parameters of the whole complex system of roads. We therefore consider 
$\tau_{-}$, $\tau_{y}$ and especially $P_{r}$ as given constraints, and explore 
the relation between the flux $f$ and the signal period $\tau$. 
Typical examples of this relation are shown in Fig. 1.

Apart from the trivial flux oscillations explained below, Fig. 1 shows a 
monotonic increase of the flux as $\tau$ grows, for the deterministic model ($\eta=0$).
For the stochastic model ($\eta>0$), however, an optimal signal period can be clearly
seen. In this case, a crossover is observed with increasing $\tau$ from the 
deterministic $f(\tau)$ to decreased values of flux. 
The optimal value of $\tau$ approaches the crossover point for small $\eta$.

\begin{figure*}
\label{1}

\centerline{
\epsfig{file=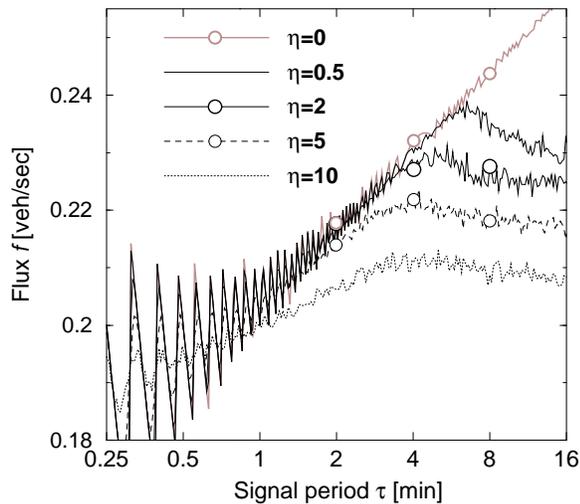, bb= 90 70 540 640, width=6.9cm, angle=270}
}
\vskip 0.3cm
\caption {Relation between signal period and flux for the values of 
acceleration noise amplitude 
$\eta=0,0.5,2,5,10m/sec^{2}$ (top to bottom). Traffic lights parameters are $P_{r} =1/3$, 
$\tau_{y} =\tau_{-} =2sec$. The total number of cars in the system is $N=400$ 
and its length is $L=10km$. The nine open circles correspond to the 
nine instances presented in Fig. 3.}

\end{figure*}

The common traffic flow and signal control theories do not predict such influence 
of $\tau$ on the flux itself (see e.g. \cite{May,Gazis,Cronje,ChangLin}). 
The average flow in each direction of a signalized intersection 
$f(\tau)$ for our consideration is expected to be 

\be
\label{flux}
f = f_0 \frac{\tau_g+\tau_y}{\tau}= f_0 \left(1-P_{r} -\frac{\tau_{-} }{\tau}\right),
\en
where $f_0$ does not depend on $\tau$ (see \cite{Gazis} and references within). 
For the signal period values $\tau\gg \tau_{-}$ that are
displayed in Fig. 1, one would expect to find $f(\tau)\approx const.$ according to 
(\ref{flux}). Therefore the numerical finding of the influence of $\tau$ on the flux 
shown in Fig. 1, revealing an optimal signal period, requires deeper investigation.

In an attempt to explain our findings, we performed extensive simulations of the
deterministic model Eq. (\ref{eqModel}) on homogeneous systems with periodic boundary 
conditions, starting from different initial conditions \cite{Sim}. Fig. 2a  presents the 
flux measured for stable stop-and-go (surface) and homogeneous (thick lines) states, showing 
our finding that the traditional 'fundamental diagram' (density-flux relation) has to be 
generalized into a three-dimensional density-wavelength-flux relation.

The projection of the surface in Fig. 2a on the density-flux plane (thin curves in Fig. 2a) 
provides a two-dimensional region in this plane, qualitatively similar to that found 
empirically by Kerner \cite{Ker1,Ker2} for synchronized flow. In our system this 
phenomena occurs since the flux of the periodic states depends not only on the density 
but also on the state's wavelength $\lambda$ \cite{FN_wavelength}. 
That is, for a given value of density, there exist many stable 
periodic states which have a range of wavelengths in the density-flux plane. Since the 
flux is also determined by the wavelength (see Fig 2b), the density-flux relation 
becomes multi-valued. 

\begin{figure*}
\label{2}

\begin{center}
 \epsfxsize=5.55cm
 \epsfbox{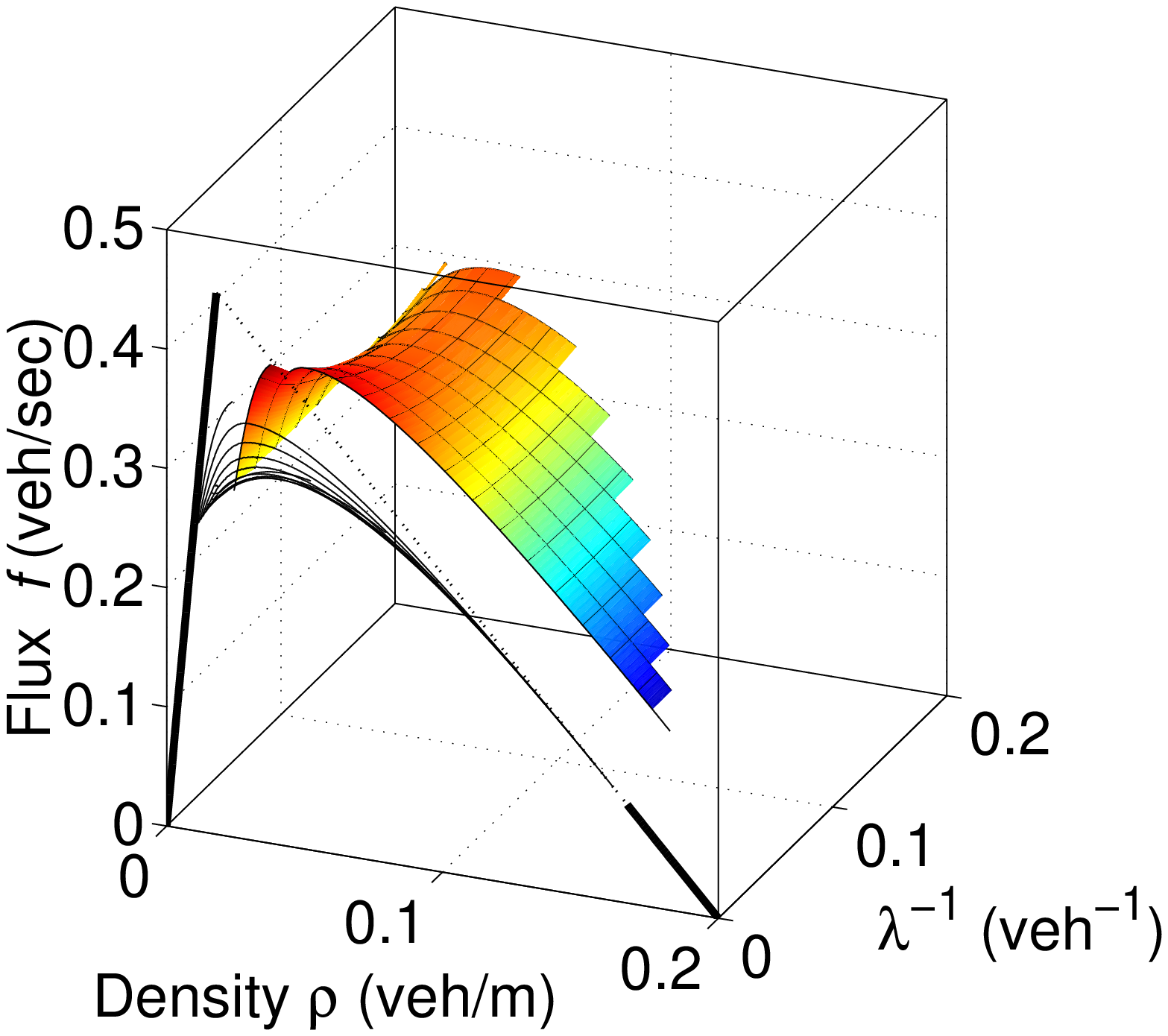}
 \epsfxsize=2.9cm
 \epsfbox{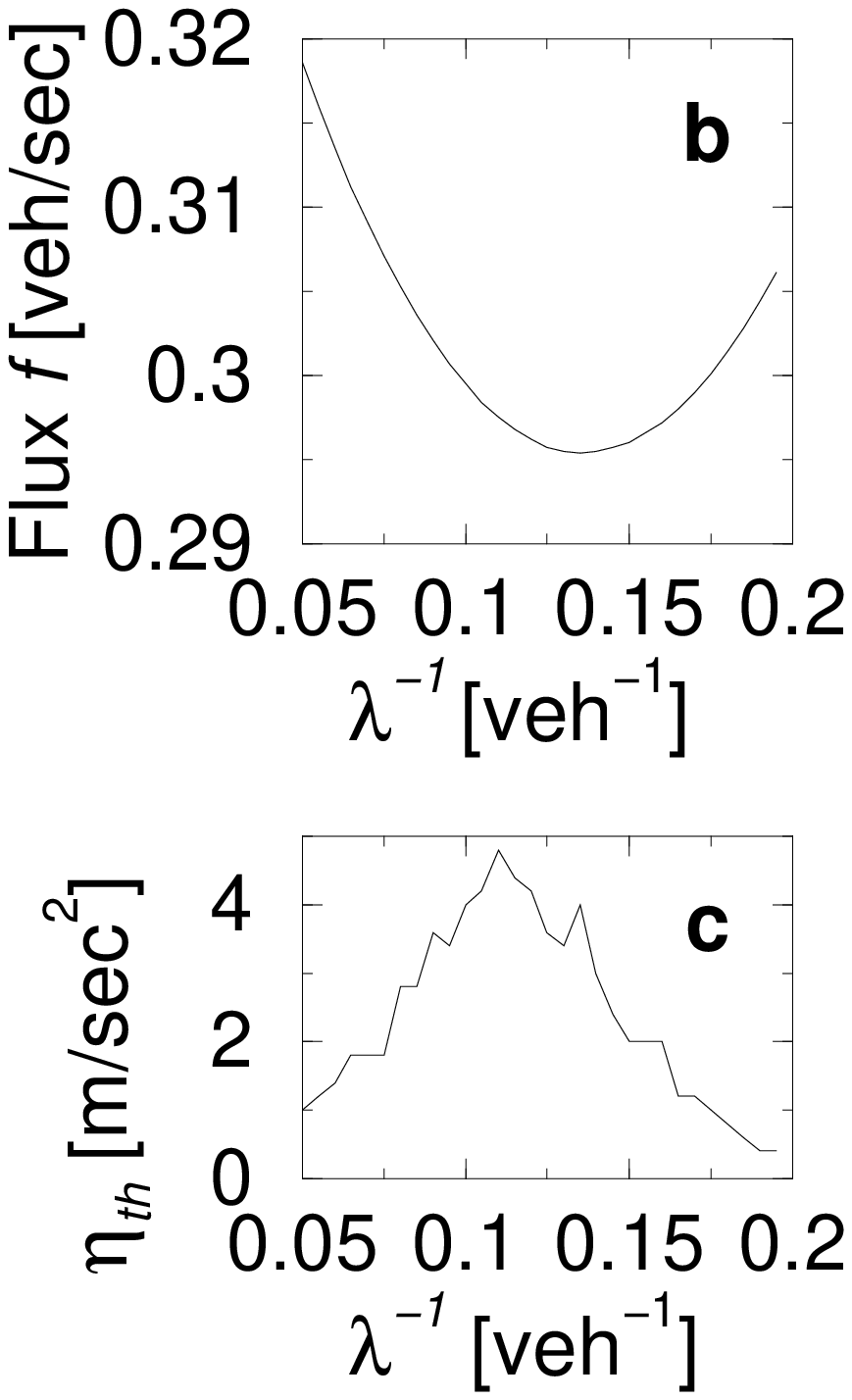}
\end{center}
\vskip 0.3cm
\caption {(a) Density-wavelength-flux relation for the different states of the 
deterministic model, including stable periodic states (surface), stable and unstable 
homogeneous states (thick and dotted lines, respectively). Some curves with fixed 
wavelength ($\lambda^{-1}=2/60, 3/60,...,12/60$) were projected on the density-flux 
plane (thin curves).
(b) A cross-section of (a) for a density $\rho=0.06veh/m$, demonstrating the typical  
dependence of the flux on the wavelength. 
(c) Noise stability threshold amplitude $\eta_{th}$, above which the states presented in 
(b) become unstable.}

\end{figure*}

Fig. 2c shows the noise stability threshold amplitude $\eta_{th}$, above which the 
states presented in Fig. 2b become unstable. A comparison of the last two figures shows 
that states with flux values close to the minimal 
are the most stable in the presence of noise, while states with higher 
values of flux (close to the upper bound of the two dimensional region) are metastable.
Therefore real-life stop-and-go traffic might be expected 
to show values of flux far below the optimum.

This new insight on the nature of the two dimensional region in the density-flux plane 
provides also an explanation to our finding of a non-trivial relation between 
flux and signal period. For $\eta=0$, the monotonically increasing $f(\tau)$ 
displayed in Fig. 1 corresponds to the left branch of Fig. 2b. 
High values of $\tau$ stimulate transition to states with high values of $\lambda$
- and therefore with high values of flux, according to Fig. 2b. These metastable states 
survive since there is no noise \cite{FN1}. The oscillations in
flux observed for relatively small $\tau$'s can be easily explained \cite{FN2}.

The crossover observed in Fig. 1 for the stochastic model ($\eta > 0$ curves) can 
be related to a crossing of the noise stability threshold $\eta_{th}$. 
For values of $\tau$ below the crossover point, the value of $f(\tau)$ is close to 
that of the 
deterministic model, since $\eta<\eta_{th}$. When $\tau$ (and therefore $\lambda$) 
is increased, two trends are expected according to Figs 2b and 2c: an increase in 
the flux and a decrease in $\eta_{th}$, until $\eta\ge\eta_{th}$, where a crossover to 
values of flux lower than that of the deterministic case. Therefore the optimal $\tau$ 
is usually close to the crossover point. Note that this crossover 
from deterministic to non-deterministic behavior with changing $\tau$ occurs in 
spite of the fact that the noise amplitude is fixed, and is related to the $\tau$ (and 
$\lambda$) dependence of $\eta_{th}$.

We therefore see that the non-trivial flux-wavelength relation is the reason 
for the unexpected behavior of $f(\tau)$. The deviations between the 
theoretical prediction of (\ref{flux}) and the numerical measurements 
presented in Fig. 1 and the crossover observed in $f(\tau)$ for $\eta > 0$ 
are related to the relation between flux and wavelength discussed above 
and to differences in noise stability threshold of different periodic states. 

To visualize the effect of the signal period and the acceleration noise amplitude on 
the flow, nine space-time diagrams are presented in Fig. 3a, below, at, and above the 
crossover. These nine diagrams correspond to the nine instances denoted by open 
circles in Fig. 1. Each dot in these space-time diagrams represents the 
position of a single car at a certain time. The dark regions therefore show the 
dense regions on the road. Looking at the diagrams in Fig. 3a, one can see the increase 
of the dominant wavelength with increasing $\tau$. But unlike the deterministic case 
where the flow is periodic and the flux is increasing with $\lambda$ and therefore with 
$\tau$, in the stochastic model ($\eta>0$) small jams emerge in the low density regions 
when the noise or the signal period exceed certain thresholds. 
In such cases, other values of wavelength, smaller than that induced by the traffic light 
are effectively involved, resulting values of flux lower than that of the $\eta=0$ case 
(see Fig. 1). 

For further support of this interpretation we evaluate the periodicity in the flow 
using single vehicle data collected at the intersection. Inspired by \cite{NSSS} 
we calculate the auto-covariance $ac_v(t)$ of the velocity function $v(t')$ measured at 
the intersection, 
\be
\label{ac}
ac_v(t)=\frac{<v(t')v(t'+t)>-<v(t')><v(t'+t)>}{<v(t')^2>-<v(t')>^2},
\en
where a linear continuation of the discrete function $v(t')$ is used. The brackets 
$<\ldots>$ indicate averaging over time $t'$. Displayed in
Fig. 3b are the auto-covariance functions for the 9 instances of Fig. 3a, respectively.
As can be seen from this figure, $ac(t=\tau)=1$ for all the $\eta=0$ instances, 
implying the flow for these cases is completely periodic, and that the time period is
equal to the period of the traffic light. For $\eta>0$, $ac_v(t=\tau)$ decreases as 
$\eta$ or $\tau$ are increased. This decrease is related to the appearance of small jams
as can be seen in Fig. 3a. The decrease in $ac_v(t=\tau)$ with $\tau$ also implies that 
noise becomes more effective as $\tau$ grows. A comparison of Fig. 3b to Fig. 1 shows 
that when $ac_v(t=\tau) \approx 1$, the flux approaches the deterministic value.

\begin{figure*}
\label{3}

\centerline{
\epsfig{file=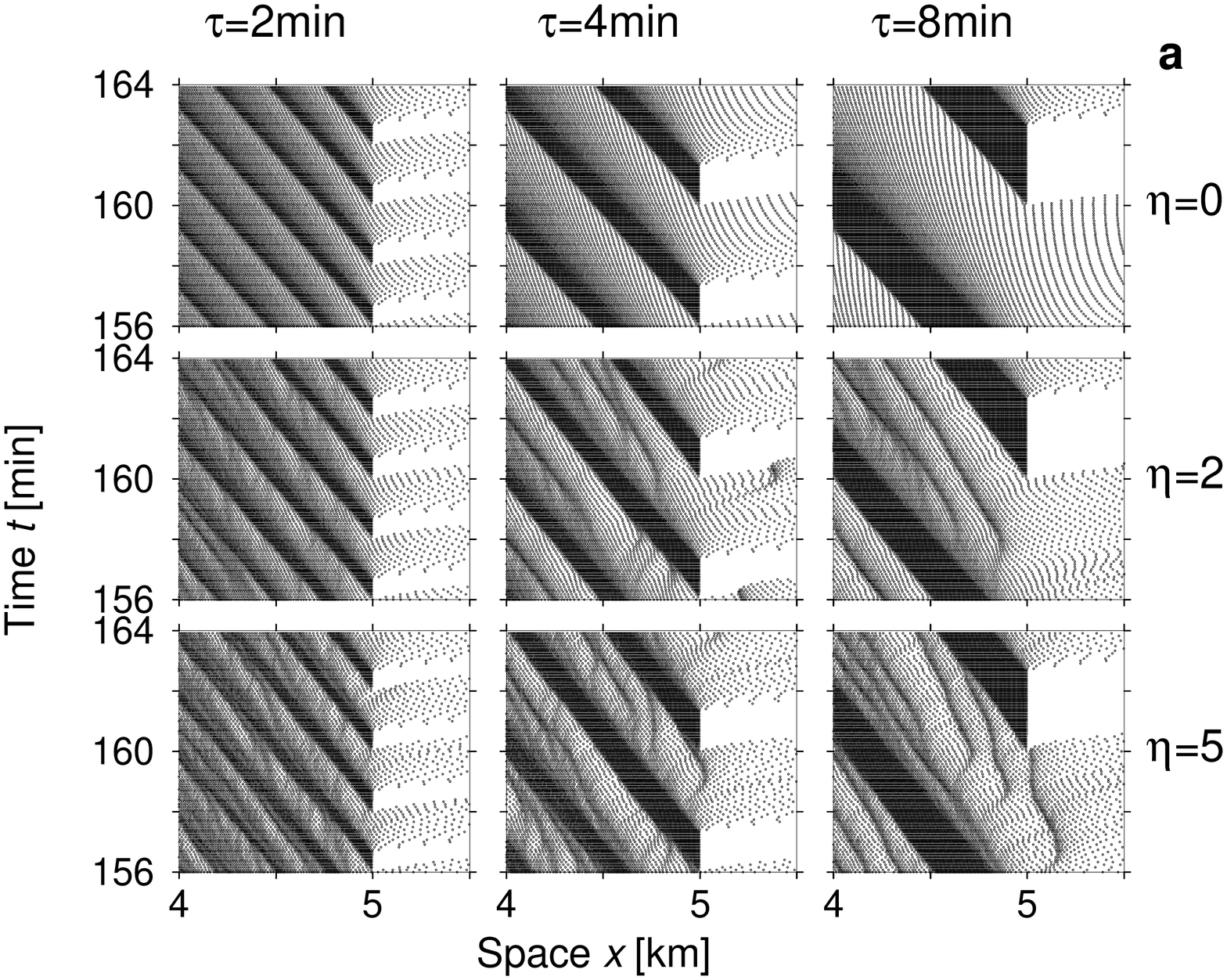, bb= 50 40 590 680, width=6.9cm, angle=270}
}
\centerline{
\epsfig{file=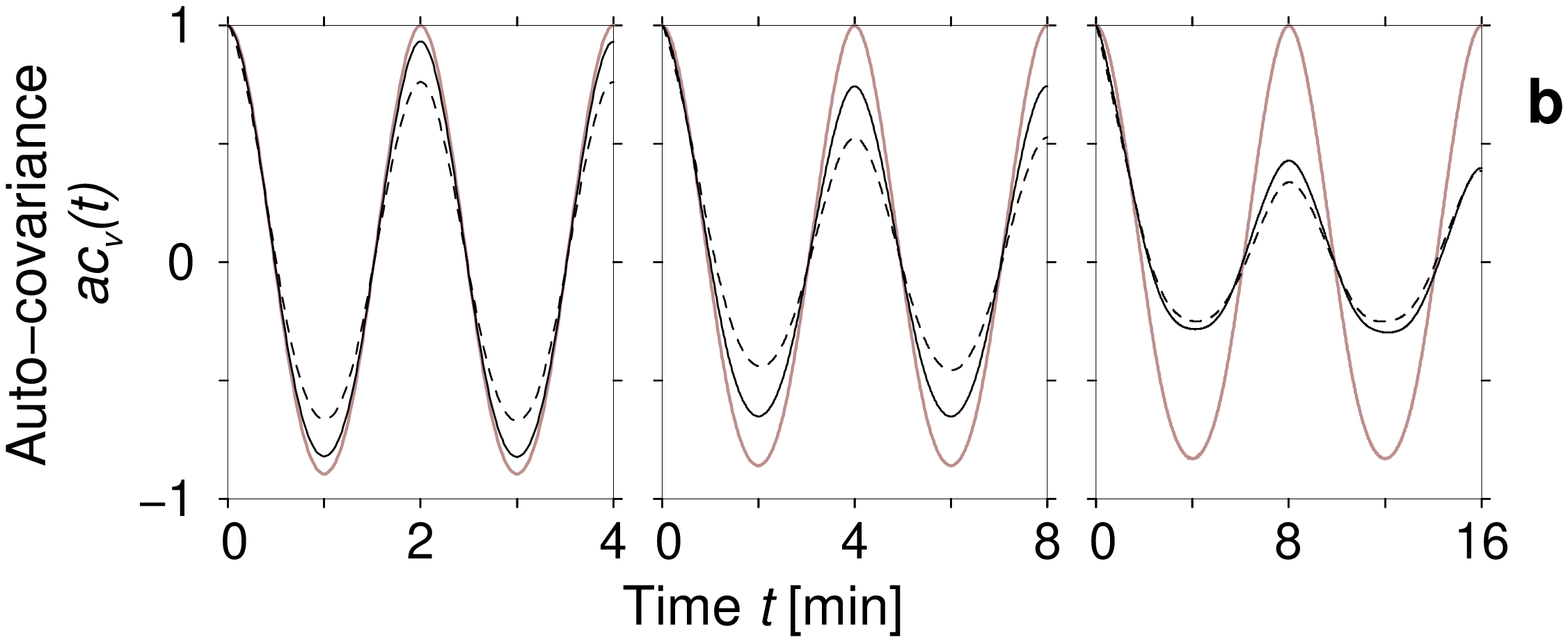, bb= 320 40 590 680, width=3.45cm, angle=270}
}
\vskip 0.3cm
\caption {(a) Space-time diagrams and (b) auto-covariance functions of systems with
single traffic light with parameters as the nine instances denoted with open circles
in Fig. 1. The position of the traffic light is at $x=5km$. Gray lines in (b) 
corresponds to $\eta=0$, solid lines to $\eta=2$, and dashed lines to $\eta=5$.}

\end{figure*}
 
{\bf (b) On-ramp with signalized entrance:}
Next, we study another example of a localized inhomogeneity, caused by 
an on-ramp. To make this inhomogeneity periodic, we introduce traffic signals at 
the downstream end of the ramp, and study its effect on the flux in the main road. 
We focus on cases where the average incoming flux $f_{in}$ is high
enough to induce congestion on the main road (see \cite{Ker0}), 
but still low enough to avoid congestion on the secondary road. Similarly to \cite{LLK}, 
we introduce also an off-ramp with equal flux of outgoing vehicles 
$f_{out}=f_{in}$ in a large distance from the on-ramp, so that the total number of cars
in the system is conserved. 
The entrance and the exit of cars from the ramps are performed in an adiabatic 
manner as in \cite{TSH1}. The incoming vehicles are allowed to enter the main
road during the green light period $\tau_{g}$, and are stopped during the red light 
period $\tau_{r}$. Here the signal period is $\tau=\tau_{g}+\tau_{r}$ and the 
relative duration of the green light is $P_{g} =\tau_{g} /\tau$. But unlike the signalized
intersection where $P_{r}$ was predetermined, here $P_{g}$ is one of the optimization 
parameters, in addition to $\tau$. The range of possible values for this parameter is
$f_{in}/f_{max}\le P_{g}\le 1$ where $f_{max}$ is the maximal possible value of the incoming 
flux. This lower bound of $P_{g}$ is considered to avoid congestion on the secondary 
road, since cars approach the queue near the traffic light with rate $f_{in}$, and this 
queue is discharged with rate $f_{max}$ during green light. 
The upper bound $P_{g}=1$ is related to an unsignalized on-ramp. 

\begin{figure*}
\label{4}
\centerline{
\epsfig{file=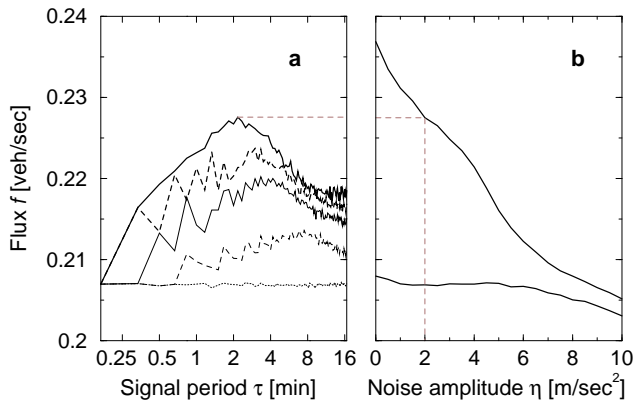, bb= 190 60 590 660, width=5.55cm, angle=270}
}
\vskip 0.3cm
\caption {(a) Relation between on-ramp signal period and flux on the main road for 
$P_{g}=0.3,0.5,0.6,0.8,1.0$ (top to bottom) and noise amplitude 
$\eta=2m/s^2$. The total number of cars in the system is $N=300$, its length is 
$L=10km$, and the flux is locally measured on the main road $100m$ upstream to 
the on-ramp. Here $f_{in}=0.1$ and $f_{max}=0.333$. (b) A comparison between the optimal flux 
(upper curve) and the flux without the presence of a traffic light (lower curve),
as a function of $\eta$.}
\end{figure*}

Typical relations between flux and signal period are shown in Fig. 4a, for different 
values of $P_{g}$ and for $\eta=2m/s^2$. The lower curve corresponds to an 
unsignalized on-ramp ($P_{g}=1$), so it is found that the introduction of a traffic light
increases the flux on the main road. Since the lower bound of $P_g$ ensures that the average influx from
the on-ramp remain the same as in the unsignalized on-ramp, we can say that the increases in the 
flux on the main road is obtained without causing congestion on the secondary road. 

As can be seen from Fig. 4a, for each value
of $P_{g}$ there usually exist a single maximum in the flux. The short relative durations
of green light $P_g \approx f_{in}/f_{max}$ usually (but not always) yields a higher flux. 
A comparison of the maximal flux at the optimal signal parameters to the flux without a 
traffic light is presented in Fig 4b, for different values of noise. The relative increase 
in the flux due to the introduction of a traffic light varies from 
$1.0\%$ for $\eta=10m/s^2$, through $10.0\%$ for $\eta=2m/s^2$ up to $13.9\%$ for $\eta=0$.

\end{multicols}


\begin{references}
\bibitem{amber}
When the light is changed from
green to amber, all simulated drivers upstream to the intersection estimate the 
intersection crossing time $t_n$ by linear
extrapolation of their position. The first car that begins to stop, $s$, is the first car that 
will not be able to cross before the light will change to red 
(i.e. $t_s=min \{ t_n|t_n>\tau_{y}  \} $). 
This car is stopped by setting $\Delta x$ in Eq. (\ref{eqModel}) to be the distance to
the traffic light position. The consecutive cars follow $s$ and stop according to 
(\ref{eqModel}). Due to this procedure cars might still pass the intersection during time 
$\tau_{y} $.
\bibitem{FN_wavelength}
The wavelength is defined as the average distance (in units of cars) between two nearest dense regions 
in stop-and-go traffic.
\bibitem{FN1}
Similar increase of the flux with decreasing $\tau$ for relatively low values of
$\tau$ is suppressed in Fig. 1 due to relatively large values of $\tau_{-}/\tau$
in (2).
\bibitem{FN2}
These oscillations are caused by the discretization of the flowing media (vehicles) 
and by the fact that all drivers are identical. Therefore, the magnitude of these 
oscillations is highest for $\eta=0$, for which the crossing times in each cycle are 
the same. It is therefore clear that the magnitude of oscillations decay linearly 
with $\tau$.
\end{references}
\end{document}